\pdfoutput=1
\documentclass[prl,superscriptaddress,nofootinbib,twocolumn,floatfix,mathbbm]{revtex4}

\usepackage{amsmath,amsfonts,graphicx,times,color,bbm}

\def\textbf#1{{\bf #1}}
\def\be{\begin{equation}}
\def\ee{\end{equation}}
\def\ben{\begin{eqnarray}}
\def\een{\end{eqnarray}}
\def\eea{\end{array}}
\def\bea{\begin{array}}

\newcommand{\bei}{\begin{itemize}}
\newcommand{\eei}{\end{itemize}}

\newcommand{\ie}{{\it{i.e.~}}}
\newcommand{\etal}{{\it{et al.}}}

\begin{document}
\bibliographystyle{apsrev}

%---------------------------------------------------------------------------

%\title{Information Causality is Inequivalent to Macroscopic Locality}
\title{Macroscopically local correlations can violate information causality}

\author{Daniel Cavalcanti}
\email{dcavalcanti@edu.nus.sg}
\affiliation{Centre for Quantum Technologies, National University of Singapore, 2 Science Drive 3, Singapore 117542}

\author{Alejo Salles}
\affiliation{Niels Bohr Institute, Blegdamsvej 17, 2100 Copenhagen, Denmark}

\author{Valerio Scarani}
\affiliation{Centre for Quantum Technologies, National University of Singapore, 2 Science Drive 3, Singapore 117542}
\affiliation{Department of Physics, National University of Singapore, 2 Science Drive 3, Singapore 117542}

\bigskip

\date{\today}

\begin{abstract}
Although quantum mechanics is a very successful theory, its foundations are still a subject of intense debate. One of the main problems is the fact that quantum mechanics is based on abstract mathematical axioms, rather than on physical principles. Quantum information theory has recently provided new ideas from which one could obtain physical axioms constraining the resulting statistics one can obtain in experiments. Information causality and macroscopic locality are two principles recently proposed to solve this problem. However none of them were proven to define the set of correlations one can observe. In this paper, we present an extension of information causality and study its consequences. It is shown that the two above-mentioned principles are inequivalent: if the correlations allowed by Nature were the ones satisfying macroscopic locality, information causality would be violated. This gives more confidence in information causality as a physical principle defining the possible correlation allowed by Nature.

%Information causality and macroscopic locality are two principles recently proposed to constrain the set of correlations allowed by Nature.
%In this paper we present an extension of information causality to arbitrary input alphabet dimensions and study its consequences. We find a scenario richer than the originally studied, which allows us in particular to show the inequivalence of both principles: if the correlations allowed by nature were the ones satisfying macroscopic locality, information causality would be violated. 
\end{abstract}

\maketitle

{\em Introduction.---}
Despite quantum theory being almost one century old, its axioms remain mathematical in nature, and we are still searching for physical principles from which to derive it. After the initial discussions of the founding fathers, the operationalist viewpoint prevailed, and most were satisfied with a useful and strikingly precise theory. With the advent of quantum information science, however, a new source from which to draw principles of a more intuitive flavor was unveiled. Following this direction, many recently works have been devoted to study which tasks would be possible if correlations more general than the ones allowed by quantum mechanics could be achieved~\cite{pr,lluis,distcomp,comcomp,ml,ic,up}. 

There are several motivations for these studies. First, they allow us to understand what is special about quantum mechanics, and consequently to get more intuition on it. Following this direction, results like the impossibility of instantaneous transmission of information \cite{pr} and of cloning, the existence of monogamy of correlation, and secrecy \cite{lluis} were shown not to be particular features of quantum mechanics. Second, if one finds a practical task limiting the correlations to be quantum, this would immediately gain the status of a physically-motivated defining property of quantum correlations. In these lines, constraints on distributed computing \cite{distcomp}, communication complexity \cite{comcomp} and the like, can be thought of as candidate requisites from which the theory could in principle be derived.  Finally, it could be possible that quantum mechanics is not the definitive description of nature, and more general correlations might be observed in the future. Thus, some physically motivated sets of correlations which are known to be bigger than the set of quantum correlations appears as candidates for possible generalizations of quantum mechanics.

A pioneering effort in this direction was that of Popescu and Rohrlich, who asked if the impossibility of transmitting information instantaneously (\ie the no-signalling principle) is enough to characterize the set of quantum correlations. It turns out that this is not the case. 
%
% introduced the no-signalling principle~\cite{pr} \daniel{Is this true???}. This, colloquially, states that one party, Alice, cannot instantaneously send messages (signal) to another distant party Bob. Although in this formulation one has to deal with the concepts of space and time, this principle can also be introduced on the level of the correlations among Alice and Bob's systems. One then states that the marginal probability of Bob getting a certain outcome upon a certain measurement choice cannot depend on Alice's measurement choice. 
%This principle, though, has proven not to be enough to characterize the set of quantum correlations. Although the correlations achievable by making local measurements on two remote (possibly entangled) quantum systems are non-signalling, 
There exist sets of correlations satisfying the no-signalling principle that are not achievable in quantum physics. The tantamount example of this is given by the Popescu-Rohrlich (PR) box~\cite{pr}.  

In this vein, other information theoretic flavored principles were proposed as candidates from which the set of physically realizable correlations could be constrained \cite{ml,ic,up}. In this work, we will be concerned with two recent proposals:  Macroscopic Locality (ML)~\cite{ml} and Information Causality (IC)~\cite{ic}. In a nutshell, ML states that the coarse-grained statistics of correlation experiments should admit a local hidden variable model, that is, those statistics do not violate any Bell's inequality~\cite{bell}. IC, on the other hand, states that if Bob receives $m$ bits of information from Alice, he cannot obtain more information about Alice's system than $m$ bits, even if he shared with her some prior resource. For the $m=0$ case, IC simply reduces to the no-signalling principle.
%IC, on the other hand, states that Bob cannot have more information about Alice's system than what Alice actually sends him. 
Formally, IC is derived by introducing a game in which Bob has to guess one of Alice's independent and random bits (chosen at random), and bounding the sum of mutual informations of Bob's different guesses with Alice's actual data by the amount of one way communication from Alice to Bob. 

The purpose of this work is twofold. First, we extend the IC game to the case in which Alice's data is composed by $d$-dimensional alphabets ({\em dits}), of which Bob has to randomly guess one, allowing in principle for $m$ dits of communication from Alice to Bob. As in the original case, we find an extremal non-signalling box that perfectly solves this task, and analyze the effect of adding noise to this set of correlations. The scenario we find in this case proves richer than the original, which allows us in turn to, second, show the inequivalence between IC and ML. In particular, we show that there exist correlations that satisfy MC, but violate IC. This implies that it is very unlikely that ML is the defining property of possibly obtainable correlations, since in this case one would have to accept that IC is violated.

We will start by reviewing the principle of ML, together with another tentative at characterizing the quantum set of correlations: the Semi Definite Program (SDP) hierarchy by Navascu\'es, Pironio, and Ac\'in (NPA)~\cite{npa}, which is intimately related with the former. We then discuss the principle of IC by introducing our task, which generalizes the original one. We then analyze the newly found scenario and compare it to the previously known case, discussing the implications. In particular, we show how, through this extension, IC comes closer to the quantum set than ML, thus gaining more %trust in it //ALEJO: no era esto lo que quería decir. era thrust nomás, pero si no les gusta ponemos otra cosa.. pero "gaining more trust in it" está mal, porque nunca está dicho quién "gains more trust"..     tal vez "momentum".. ?
thrust as a candidate for a physical axiom defining the set of quantum correlations.

\bigskip
\emph{Macroscopic Locality and the NPA hierarchy.---}
Macroscopic Locality~\cite{ml} was introduced more as a principle producing an alternative to quantum theory than as an axiom for it. Since its inception, it was known that, when considered together with the no-signalling principle, it does not give rise to the set of quantum correlations, but, instead, to a larger set. This set is precisely the one labeled by $Q_1$ in~\cite{npa}, that is, the first step in a hierarchy of SDPs that eventually converge to the quantum set. It was known already that, in some scenarios, further steps in the hierarchy are strictly contained in $Q_1$, while also containing the quantum set~\cite{npa}. Thus, it sufficed to prove the equivalence of $Q_1$ and the ML set to realize that this axiom would not suffice to fully determine quantum correlations. 

ML retains its interest, though, both as an example of principle with physical content that can be put forth as candidate for axiom, and as a testable requirement from which a theory larger than quantum arises, and which could in principle be used to disprove quantum physics. As we mentioned before, it demands that classical physics be recovered in the large particle number limit by imposing that the coarse-grained correlations arising in an experiment involving a macroscopic number of particles can be modeled with local hidden variables. This requirement applied to microscopic probabilities is what gives Bell local sets of correlations, which satisfy Bell inequalities. Less restraining, ML imposes this constraint on their macroscopic counterpart; all Bell local correlations thus satisfy ML, but the converse fails to hold. To date, ML had not been contrasted with other attempts at axiomatizing quantum correlations from physical postulates.

\bigskip
\emph{Information Causality and its extension to arbitrary alphabet dimensions.---}
As we stated before, Information Causality was introduced through the use of a game with two players, Alice and Bob, and in which Alice is given $N$ independent and random bits of which Bob has to guess one, chosen at random. In order to achieve this, they can share prior to the game any amount of physical resources they want (classical correlations, entangled states, or, in a hypothetical scenario, nonlocal post-quantum boxes), and Alice is further allowed to send $m$ bits to Bob after she receives her input. We generalize this task by extending the alphabet in which Alice's inputs range from 2 to an arbitrary number $d$ of dimensions. Bob then has to guess, again using prior shared resources but now allowing Alice to send him $m$ {\em dits}, one of Alice's inputs, again chosen at random. 

%\begin{figure}[htbp]
%\includegraphics[width=0.5\textwidth]{Figures/ICTaskSketch}
%\caption{Information Causality game for $d$-dimensional alphabets.}
%\label{fig:ICTask}
%\end{figure}

IC, in fact, imposes a bound on
\begin{equation}
I\equiv \sum_{K=0}^{N-1} I(x_K:G|y=K),
\end{equation}
where %according to the notation introduced in Fig.~\ref{fig:ICTask}, 
$x_K$ are Alice's independent and random input dits, $y$ is Bob's random input telling him which $x_K$ to aim for, and $G$ is Bob's guess at that chunk of Alice's data. $I(x_K:G|y=K)$ is in turn the Shannon mutual information between Bob's guess and Alice's input $x_K$, given that he aims at guessing that particular input. IC, stated in general, imposes that:
\begin{equation}
I\leq m \log_2 d.
\label{eq:IC}
\end{equation}
While all quantum correlations satisfy this requirement, it is still an open question whether there are post-quantum correlations that satisfy IC~\cite{allcock}.

\begin{figure}[htbp]
\includegraphics[width=0.45\textwidth]{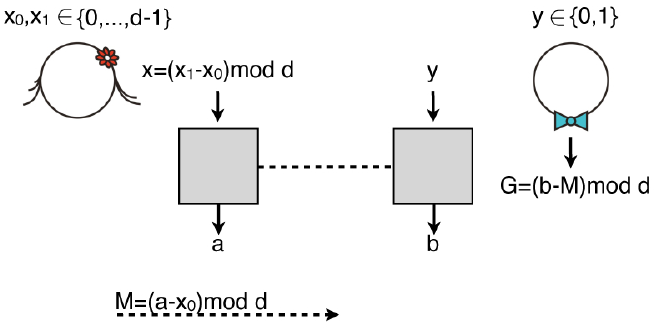}
\caption{\textbf{Information Causality protocol for $d$-dimensional alphabets.} Suppose Alice is given two dits, $x_0$ and $x_1$, while Bob gets a bit $y$. Bob is asked to guess the value of one of Alice's dits, according to the value of $y$. In the case that Alice and Bob share the noiseless box~\eqref{eq:PR} they can solve this problem perfectly. In order to do that Alice inputs $(x_1-x_0)\mod d$ into the box, while Bob inputs $y$. After receiving her output $a$, Alice sends a message $M=(a-x_0)\mod d$, corresponding to one dit of communication. Bob, in possession of $M$, makes his guess $G=(b-M)\mod d=(b-a+x_0)\mod d$. Given that the box behaves as $x.y=(b-a)\mod d$, Bob computes the value $G=[(x_1-x_0).y+x_0]\mod d$, which equals $x_0$ if $y=0$ and $x_1$ if $y=1$. }
\label{fig:ICProtocol}
\end{figure}

We now study to what extent the IC condition defines the quantum set. In order to do this, we consider a protocol which perfectly solves the task when an extremal set of non-signalling correlations is used as a resource. By adding noise to this box, we approach the quantum set. We start by considering the case $N=2$. Alice and Bob share a box with $d$ inputs on Alice's side and 2 inputs on Bob's, and $d$ outputs on both sides (see Fig.~\ref{fig:ICProtocol}). We label by $a$ ($b$) the  output of the box on Alice's (Bob's) side. The message Alice sends to Bob is $M=(a-x_0)\mod d$, and Bob's guess is $G=(b-M) \mod d = (b-a+x_0)\mod d$. 

The probability of Bob making a correct guess is:
\begin{equation}
\begin{split}
&P_S=P(G=x_0, y=0)+ P(G=x_1,y=1)=\\
&\frac{1}{2} [ P(G=x_0 | y=0) + P(G=x_1 | y=1) ]=\\
&\frac{1}{2 d} \sum_{j=0}^{d-1} [ P(G =x_0 | x_0=j, y=0) + P(G = x_1 | x_1=j, y=1) ],
\end{split}
\end{equation}
where we used that both Alice and Bob's inputs are unbiased. In the case $y=0$, the condition for a correct guess $G=x_0$ is equivalent to $b-a = 0 \mod d$. When $y=1$, we need to split the cases according to $x=j$ with $j=0,\ldots,d-1$, and the correct guess condition $G=x_j$ is equivalent, in each case, to $b-a =j \mod d$. Putting this together, we have for the success probability:
\begin{equation}
\begin{split}
P_S=\frac{1}{2 d} \sum_{j=0}^{d-1} [ &P(b-a = 0 \mod d | x_0=j, y=0) + \\
&P(b-a = j \mod d | x_1=j, y=1) ].
\end{split}
\end{equation}

Consider now the case in which Alice and Bob share a box defined as:
\begin{equation}
{\rm PR}_{0}(ab | xy) \equiv \left\{ \begin{array}{ll}
1/d & {\rm if}\; x.y=(b-a) \mod d \\
0 & {\rm otherwise}
\end{array} \right.,
\label{eq:PR}
\end{equation}
with $a, b, x \in \{0, \ldots, d-1\}$, and $y \in \{0,1\}$. This is a straightforward generalization of the PR box~\cite{pr}, to which it reduces in the case $d=2$. It is extremal in the $d2dd$ non-signalling polytope, as can be verified by using the same counting argument as in the proof of Theorem 1 of Ref.~\cite{barrett}. This box clearly satisfies $P(b-a = x.y \mod d |xy) = 1$, which leads to a probability of success equal to unity (which is the case depicted in Fig.~\ref{fig:ICProtocol}). This in turn implies $I=2 \log_2 d$, which violates the IC condition~\eqref{eq:IC} since $m=1$.

We now define an isotropic noisy box as follows:
\begin{equation}\label{eq:PRN}
\begin{split}
{\rm PR}(E) \equiv &E \; {\rm PR}_0 + (1-E) {\mathbbm 1}\\
=& E \; {\rm PR}_0 + \frac{1-E}{d}({\rm PR}_0+\ldots+{\rm PR}_{d-1}),
\end{split}
\end{equation}
that is, a mixture of the generalized PR box~\eqref{eq:PR} and a box $\mathbbm 1$, representing classical random noise. Box $\mathbbm 1$ outputs completely random dits regardless the inputs, and can be decomposed, as we did in the second line of Eq.~\eqref{eq:PRN}, in terms of ${\rm PR}_0$ plus $d-1$ other extremal non-signalling boxes, denoted by ${\rm PR}_j$ with $j\in \{1,\ldots,d-1\}$, s.t.
\begin{equation}
{\rm PR}_{j}(ab | xy) \equiv \left\{ \begin{array}{ll}
1/d & {\rm if}\; x.y=(b-a)+j \mod d \\
0 & {\rm otherwise}
\end{array} \right..
\label{eq:PRj}
\end{equation}
%which are obtained from ${\rm PR}_0$ by making cyclic permutations of the outcomes. 
The parameter $E$ ($0 \leq E \leq 1$) quantifies the amount of noise in the box. This noisy correlations satisfy $P(B-A = x.y \mod d |xy) = \frac{(d-1)E+1}{d}$ so we have for the success probability $P_S=\frac{(d-1)E+1}{d}$.

Although we could search for the critical value of $E$ for a single box to stop violating IC, it was seen already in the original $d=2$ case that this is not optimal~\cite{ic}. Instead, one needs to consider the task illustrated in Fig.~\ref{fig:ICProtocol} for an arbitrary input size $N$, and extend the protocol described by nesting many instances of it.
% \daniel{YO SACARIA ESO: This highlights a crucial feature of the principle of IC: even if one set of correlations might seem to satisfy it, this could always be due to the suboptimality of the protocol considered. That is, there could in principle exist a better protocol that exploits the correlations in a deeper way, yielding a violation of IC. Hence, when considering the extent of the set defined by IC, we are limited to dealing with outer bounds, since a better protocol could always reduce its size (except of course if an explicit proof of IC compliance is given, as in the quantum case). DE ACUERDO CON SACAR - Valerio}

For the case of isotropic noisy boxes and $d=2$, it was seen that a suitable nesting of the protocol was enough to recover the quantum bound~\cite{ic}. In the Appendix, we describe this nesting and its extension to arbitrary dimension $d$. This extension proves straightforward, apart from the consideration of the cancellation of errors of different boxes, which now happens less frequently due to the larger outcome alphabet. We find a recurrence relation for the probability of success, and by solving it, we get for the success probability of $n$ boxes (see the Appendix for details):
\begin{equation}
P_S^{(n)}=\frac{(d-1) E^n+1}{d}.
\label{eq:PSN}
\end{equation} 
This is the probability of Bob producing a correct guess after the use of $n$ boxes. Note that, given the structure of the nesting, one needs to use $n=\log_2 N$ boxes if the input consists of $N$ dits. 

In order to study the violation of IC in terms of the probability of a successful guess, we use Fano's inequality~\cite{fano} to bound $I$, as was already discussed for arbitrary $d$-dimensional alphabets in~\cite{ic}:
\begin{equation}
I\geq 2^n \left( \log_2 d -  h(P_S) -  (1-P_S) \log_2 (d-1) \right),
\end{equation}
where $h(p)= -p \log_2 p - (1-p) \log_2 (1-p)$ is the binary entropy, and we set equal probabilities for the correct guessing of the different dits, as is the case in the protocol. Considering then this bound for the success probability in the nested protocol given in Eq.~\eqref{eq:PSN}, we obtain:
\begin{equation}
\begin{split}
I \geq 2^n &\left[ \log_2 d - h\left( \frac{1+(d-1) E^n}{d} \right) +\right.\\
&\left. \quad \frac{d-1}{d} (1-E^n) \log_2(d-1) \right].
\end{split}
\label{eq:fanoBound}
\end{equation}

\begin{figure}[htbp]
\includegraphics[width=0.5\textwidth]{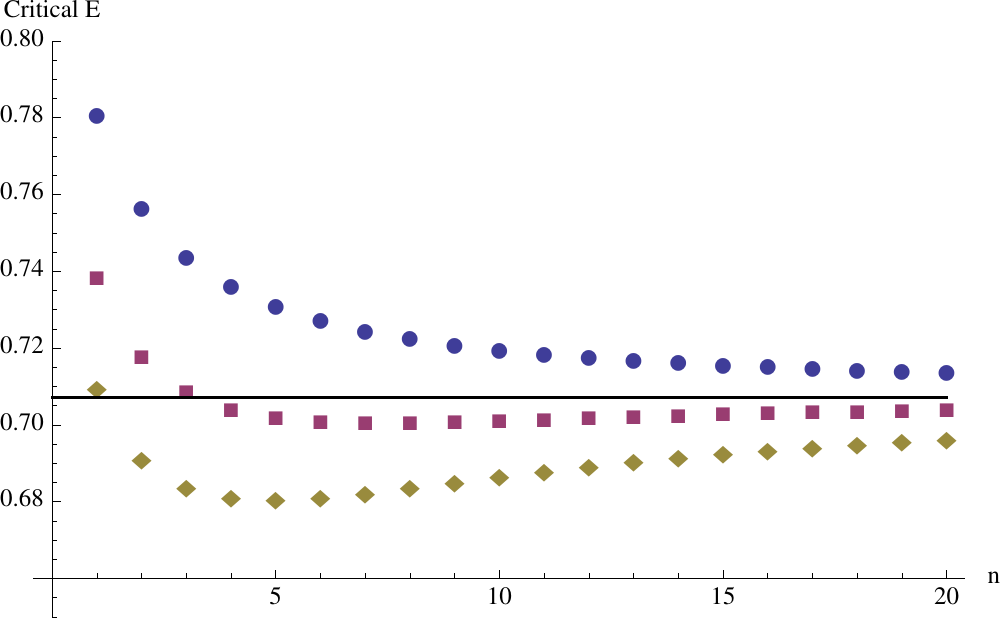}
\caption{{\bf Critical noise level for IC violation.} We plot the critical amount of noise for which IC ceases to be violated, as a function of $n$, and for different values of $d$ ($d=2$ circles, $d=5$ squares, $d=10$ diamonds). The solid line corresponds to $1/\sqrt{2}$, which, according to our numerical calculations up to $d=5$, coincides with ML.} 
\label{fig:criticalNoise}
\end{figure}

\bigskip
{\em Consequences of the IC extension and inequivalence with ML.---}
We now turn to the consequences of the extension of IC we have introduced. In Fig.~\ref{fig:criticalNoise} we show the critical value of $E$ for which IC ceases to be violated, as a function of the number of boxes used, $n$, and for different values of $d$. This value is obtained from~\eqref{eq:fanoBound}. For $d=2$, the critical value of $E$ asymptotically approaches Tsirelson's bound $E_T=1/\sqrt{2}\simeq 0.707$, which defines the extent of the quantum set for isotropic correlations. Therefore, we recover the result that IC defines isotropic quantum correlations for $d=2$, which also coincide with those satisfying ML (or, equivalently, are in $Q_1$). It was shown, however, that this protocol does not define the entire quantum boundary even for $d=2$~\cite{allcock}.

For larger values of $d$, however, the situation proves richer. First, we notice that for sufficiently large $d$, the curves go below the critical value $E_T$. This value is therefore no longer representative of the extent of quantum correlations. However, it still defines the extent of Macroscopically Local isotropic correlations, as we have checked for values of $d$ up to 5 by solving the SDP that defines the set $Q_1$. This thus proves the inequivalence of ML and IC. Table~\ref{tab:results} summarizes the critical values obtained for the different sets.

\begin{table}[htbp]
\begin{tabular}{|c|c|c|c|}
\hline
$d$ & $E_{\rm IC}$ & $E_{\rm ML}$ & $E_{\rm Q}$\\
\hline
 2 & 0.707 & 0.707 & 0.707\\
 3 & $\lesssim$ 0.708 & 0.707 & ? \\
 4 & $\lesssim$ 0.705 & 0.707 & ? \\
 5 & $\lesssim$ 0.700 & 0.707 & ? \\
\hline
\end{tabular}
\caption{{\bf Critical values of noise for increasing alphabet dimension.} Here, $E_{\rm IC}$ is the critical noise for which IC ceases to be violated using the protocol described, optimized over $n$. $E_{\rm ML}$ is the corresponding value for ML correlations ($Q_1$), and $E_{\rm Q}$ is the extent of the quantum set, ignored apart from the $d=2$ case. All values to $10^{-3}$ accuracy.}
\label{tab:results}
\end{table}

\begin{figure}[htbp]
\includegraphics[width=0.45\textwidth]{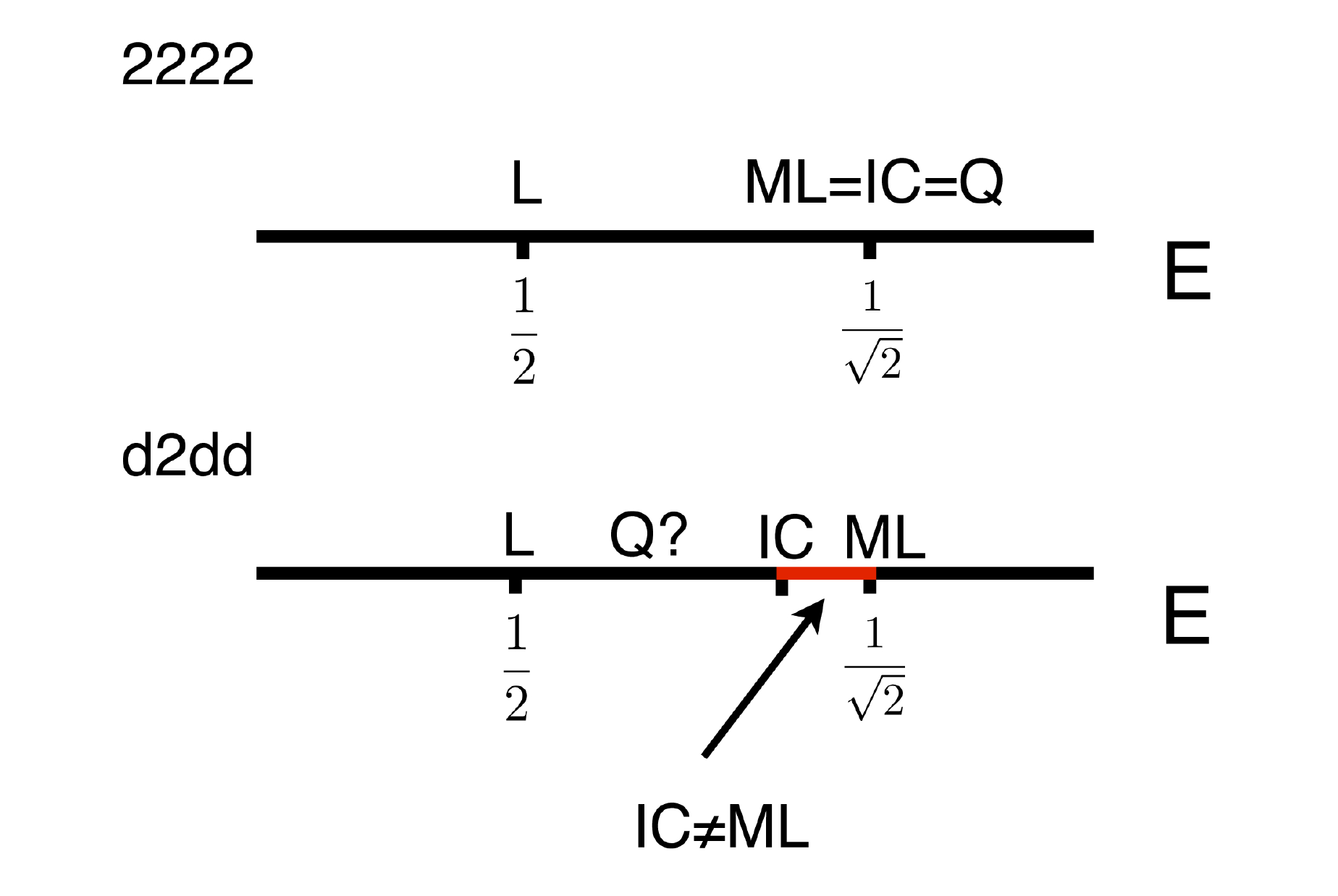}
\caption{\textbf{Scenario comparison for the $2222$ and $d2dd$ cases.} In the $2222$ case both IC and ML stop being violated for the same amount of noise $E=1/\sqrt{2}$. For the case $d2dd$ with $d>2$, the violation of IC can stand more noise than that of ML, which shows that ML correlations can violate IC.}
\label{fig:comparison}
\end{figure}

Unfortunately, using either numerical searches or plausible analytical guesses, we were unable to find quantum realizations for the isotropic correlations defined by $E_{\rm IC}$ for $d > 2$ above the local boundary. So, contrary to the $d=2$ case, we do not know if the IC strategy studied here reaches the quantum boundary.
% \textbf{THIS FIGURE IS NOT REALLY NEEDED.} ALEJO: la incluimos (con líneas) de nuevo, así que resucito el comentario que sigue
We sketch the situation pictorially in Fig.~\ref{fig:comparison}, where we compare the $d=2$ and larger $d$ cases.

Finally, we stress a curious feature of the curves presented in Fig.~\ref{fig:criticalNoise}. Differently from the $d=2$ case, in which the critical level of noise monotonously approaches $E_T$, for larger values of $d$ the curves display a minimum in the critical noise for a finite value of $n$. The position of this minimum further decreases with increasing $d$. While this effect is not yet fully understood, it might hint towards the non-optimality of the protocol considered. We leave this question for further investigations.

\bigskip
{\em Discussion.---} We have extended the principle of IC to arbitrary input alphabet dimensions and studied the consequences of this extension. The scenario we found turned out to be richer than the original case, even if only considering isotropic correlations. This allowed us to show the inequivalence between IC and ML. The extension we presented might prove useful in deciding whether IC is a good candidate as axiom for quantum correlations, question that still remains open. 

As a final remark, we note that the extension remains valid in the $d \to \infty$ case. Interestingly, through this extension, Bob would be able to guess (to arbitrary precision) either one of two arbitrary precision floating point numbers of Alice using just one of the extremal $d2dd$ boxes we introduced. How this compares as a resource to an asymptotic number of standard PR $2222$ boxes is a question for further study.

\bigskip
{\em Acknowledgements.---} We would like to thank Tomasz Paterek, Nicolas Brunner, Andreas Winter and Michael Wolf for discussions. We acknowledge financial support by the EU STREP COQUIT under FET-Open grant number 2333747, National Research Foundation and the Ministry of Education of Singapore.

%\end{document}

\section{Appendix: Nested protocol}
%\label{sec:methods}

\begin{figure*}[htbp]
\includegraphics[width=0.6\textwidth]{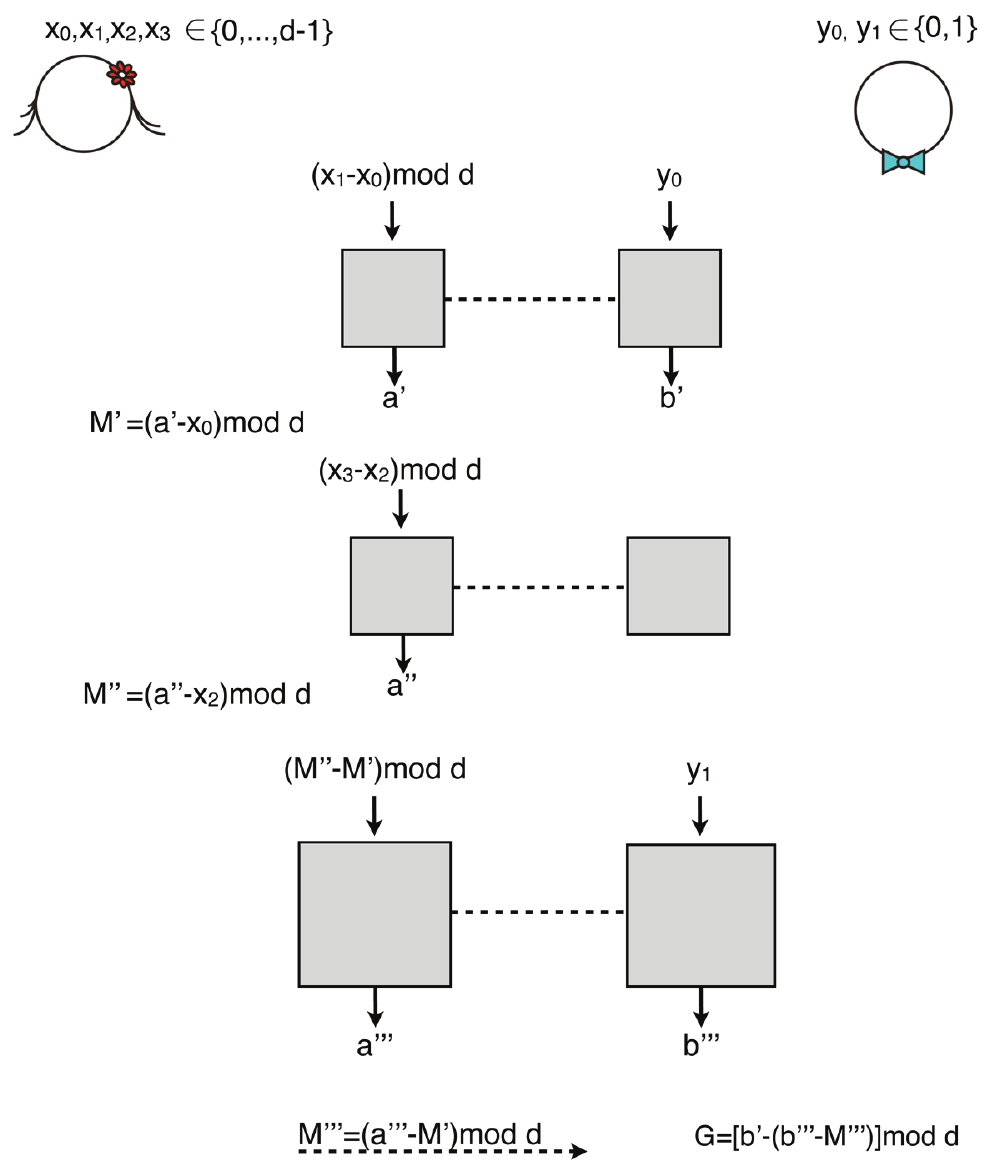}
\caption{{\bf Nesting of the protocol for $N>2$}. We illustrate how to recursively employ the $N=2$ protocol in order to solve the IC task for larger values of $N$. Depicted is the case of $N=4$, where Bob aims to know the value of either of the four dits $x_0, x_1, x_2,$ or $x_3$. Alice inputs $(x_1-x_0)\mod d$ in the first box, and would have to send the message $M'=(a'-x_0)\mod d$ to make Bob to be able to know either $x_0$ or $x_1$. She proceeds similarly in the second box, but now using $x_2$ and $x_3$. In this case the message she would send would be $M''=(a''-x_2)\mod d$. Since she can send only one message, they use a third box to make Bob able to know either $M'$ or $M''$. To this end, Alice inputs $(M''-M')\mod d$ to the third box, and sends the message $M'''=(M''-M')\mod d$ to Bob. By inputing $y_1=0,1$ to the third Box he can guess the value of $M'$ or $M''$, depending on which dit he is looking for. Having this information he uses either box 1 or box 2 to discover the dit he aims for. In the figure we supposed that Bob wants to know the value of either $x_0$ or $x_1$. In this case he would input $y_1=0$ in the third box, find the value of $M'$, input $y_0$ to first box and finally guess the value of the bit he wants according to $G=[b'-(b'''-M''')]\mod d$. For arbitrary $N$ this process can be iterated straightforwardly.}
\label{fig:Nesting}
\end{figure*}

Here we discuss the nesting of the protocol described in the main text needed in order to tackle the task when $N>2$. The nesting process is similar to that originally presented in~\cite{ic}, but differs in the treatment of accumulated errors. For clarity, suppose Alice is given $N=4$ dits, $x_0, x_1, x_2,$ and $x_3$, of which Bob has to learn one, indexed by the random bits he receives, $y_0$ and $y_1$ (see Fig.~\ref{fig:Nesting}). Now Alice and Bob share three boxes. If these were the noiseless ${\rm PR}_0$ boxes of Eq.~\eqref{eq:PR}, they could perfectly solve the task. The main idea (explained in details in Fig.~\ref{fig:Nesting}) is that they can the first box to reveal the value of either $x_0$ or $x_1$, the second box to $x_2$ or $x_3$. In the first case Bob would have to know a message $M'$, while in the second $M''$. Since only one message can be transmitted, they use the third box to reveal to Bob either $M'$ or $M''$, depending on which dit he wants to know.

%Bob chooses to input $y_0$ to one of the two first boxes, according to the value of $y_1$. With the value of $M'$ and $M''$ in hand, he would be then able to guess the desired dit, as in the simple protocol. These two values are then encoded by use of the third box, the outcome of which is included in Alice's classical message to Bob. For instance, suppose that Bob wants to learn the value of $x_0$. So he inputs $y_1=0$ in the third box chooses the first box. By inputting $y_1$ to the third box, he learns the value of $M'$, with which he finally reconstructs the desired value by combining it with $b'$.

Clearly, this process can be repeated an arbitrary amount of times, with Alice and Bob sharing $2^n-1$ boxes for a task involving $N=2^n$ dits. Note however that only $n$ of these boxes are actively used in the protocol, in the sense that the outcome of the remaining ones is irrelevant for the correctness of the guess. So now we look at how the errors accumulate in the different stages of the nested protocol. The joint probability of success for two different boxes $\alpha$ and $\beta$ is given by:
\begin{equation}
P_S^{\alpha\beta}=P_S^\alpha P_S^\beta + \frac{1}{d-1} (1-P_S^\alpha) (1-P_S^\beta),
\end{equation}
where the first term corresponds to both boxes succeeding, and the second term arises from error compensation: of the $(d-1)^2$ ``$\alpha$ wrong, $\beta$ wrong'' events, only $d-1$ lead to a cancellation. We can now think of the probability of success for $n+1$ boxes as the joint success of $n$ boxes and a single box:
\begin{equation}
P_S^{(n+1)}=P_S^{(n)} P_S + \frac{1}{d-1} (1-P_S^{(n)})(1-P_S).
\end{equation}
This recurrence can be solved for $P_S=\frac{(d-1) E + 1}{d}$, the probability of success of a noisy PR box, to finally get:
\begin{equation}
P_S^{(n)}=\frac{(d-1) E^n +1}{d}.
\end{equation}


\begin{thebibliography}{30}

\bibitem{pr} S. Popescu, D. Rohrlich, Found. Phys. {\bf 24}, 1379 (1994).

\bibitem{lluis} L. Masanes, A. Ac\'in, and N. Gisin, Phys. Rev. A. {\bf 73}, 012112 (2006).

\bibitem{distcomp} N. Linden, S. Popescu, A. J. Short, and A. Winter, Phys. Rev. Lett. {\bf 99}, 180502 (2007).

\bibitem{comcomp}W. van Dam, arXiv quant-ph/0501159 (2005). G. Brassard, \etal, \prl {\bf 96}, 250401 (2006).

\bibitem{ml}  M. Navascu\'es and H. Wunderlich, Proc. Roy. Soc. Lond. A  {\bf 466} 881-890 (2009).

\bibitem{ic} M. Paw\l owski, T. Paterek, D. Kaszlikowski, V. Scarani, A. Winter, M. \.Zukowski, Nature {\bf 461}, 1101-1104 (2009).

\bibitem{up}J. Oppenheim, S. Wehner, arXiv:1004.2507v1.

\bibitem{bell} J. S. Bell, Physics (Long Island City, N.Y.) {\bf 1}, 195 (1964).

\bibitem{npa} M. Navascu\'es, S. Pironio, A. Ac\'in, Phys. Rev. Lett. {\bf 98}, 010401 (2007); New J. Phys. {\bf 10}, 073013 (2008).

\bibitem{allcock} J. Allcock, N. Brunner, M. Paw\l owski, V. Scarani, Phys. Rev. A {\bf 80}, 040103 (2009).

\bibitem{fano} M. Nielsen, I. Chuang, {\em Quantum Information and Computation} (Cambridge University Press, Cambridge, UK, 2001).

\bibitem{barrett}J. Barrett, N. Linden, S. Massar, S. Pironio, S. Popescu, D. Roberts, Phys. Rev. A {\bf 71}, 022101 (2005).



\end{thebibliography}
\end{document}